\documentclass[prl,preprintnumbers,amsmath,amssymb,tightenlines,twocolumn,superscriptaddress]{revtex4}
\usepackage[utf8]{inputenc}
\usepackage{varioref}
\usepackage{units}
\usepackage{graphicx}
\DeclareGraphicsExtensions{.eps}
\usepackage{dcolumn}
\usepackage{bm}
\usepackage{epsfig}
\usepackage{stmaryrd}
\usepackage{ifthen}

\newcommand{\ssection}[1]{{\noi  \it #1:}}

\def\noi{\noindent}
\def\beq{\begin{equation}}
\def\eeq{\end{equation}}

\newcommand{\eref}[1]{Eq.~(\ref{#1})}

\newcommand{\cref}[1]{chapter~\ref{#1}}

\newcommand{\Cref}[1]{Chapter~\ref{#1}}

\newcommand{\ii}{\mathrm{i}}

\newcommand{\ZZ}{\cc{z}}
\newcommand{\adjZZ}{\mathrm{D}}
\newcommand{\Reals}{\mathbb{R}}
\newcommand{\unity}{\openone}
\newcommand{\ee}{\vec{e}}
\newcommand{\kk}{\vec{k}}

\newcommand{\Tr}{\operatorname{Tr}}

\newcommand{\cc}[1]{{#1}^*}
\newcommand{\abs}[1]{\left\vert #1 \right\vert}
\newcommand{\op}[1]{#1}
\newcommand{\adj}[1]{\op{#1}^\dagger}
\newcommand{\E}[1][\empty]{
  \ifthenelse{\equal{#1}{\empty}}
    {\mathbb{E}}
    {\mathbb{E}\left( #1 \right)}
}
\renewcommand{\exp}[1][\empty]{
  \ifthenelse{\equal{#1}{\empty}}
    {\mathrm{exp}}
    {\mathrm{e}^{#1}}
}
\newcommand{\intl}[3]{\int_{#1}^{#2}\mathrm{d}#3\,}
\renewcommand{\vec}[1]{\boldsymbol{#1}}
\newcommand{\bra}[1]{\langle #1 |}
\newcommand{\ket}[1]{| #1 \rangle}
\renewcommand{\sp}[2]{\langle #1 | #2 \rangle}
\newcommand{\BathCor}{\alpha}
\newcommand{\K}{\mathcal{K}}
\newcommand{\kB}{}
\newcommand{\blankout}[1]{\null}
\usepackage{ulem}  
\usepackage{color}

\definecolor{darkgreen}{RGB}{0,128,0}

\newcommand{\quotes}[1]{``#1''}

\begin{document}

\title{Hierarchy of stochastic pure states for open quantum system dynamics}
\author{D.~S\"u\ss}
\altaffiliation[Present address: ]{Centre for Quantum Dynamics, Griffith University, Brisbane 4111 Australia}
\affiliation{Institut f\"ur Theoretische Physik, Technische Universit\"at Dresden,D-01062 Dresden, Germany}

\author{A.~Eisfeld}
\email{eisfeld@pks.mpg.de}
\affiliation{Max Planck Institute for the Physics of Complex Systems, N\"othnitzer Strasse 38, 01187 Dresden, Germany}
\author{W.~T.~Strunz}
\affiliation{Institut f\"ur Theoretische Physik, Technische Universit\"at Dresden,D-01062 Dresden, Germany}

\begin{abstract}
We  derive a hierarchy of stochastic evolution equations for  pure states (quantum trajectories) to efficiently solve open quantum system dynamics with non-Markovian structured environments.  
From this {\em hierarchy of pure states} (HOPS) the exact reduced density operator is obtained  as an ensemble average.
We demonstrate the power of HOPS by applying it to the Spin-Boson model, the calculation of absorption spectra of molecular aggregates and energy transfer in a photosynthetic pigment-protein complex.

\end{abstract}

\maketitle

The treatment of the dynamics of realistic open quantum systems still poses both conceptual and computational challenges.
These arise from non-Markovian behavior due to a structured environment or strong system-environment interaction \cite{We08__,MaKue00__}.
Severe assumptions,  like weak-coupling  or Markov approximation, are often made for practical reasons.
 However, they fail for many  systems of interest.
In these situations one relies on computationally demanding numerical methods.
Among these are path integral approaches \cite{MaMa95_quapi,ThReHa00_quapi} or hierarchical equations of motion \cite{Ta06_082001_,KrKrRo11_2166_} for the system's reduced density matrix.

In this Letter we follow a different strategy and derive a hierarchy of stochastic differential equations for  pure states in the system Hilbert space (quantum trajectories).
From this {\em hierarchy of pure states} (HOPS) the exact reduced density operator is obtained  as an ensemble average.
Our approach is based upon \emph{non-Markovian Quantum State diffusion} (NMQSD),
 derived in its general form in Refs.~\cite{St96_25_,Di96_309_,DiSt97_569_,DiGiSt98_1699_}.
NMQSD has been applied to various physical problems including the  description of energy transfer in photosynthesis \cite{RiRoSt11_113034_,RiRoSt11_2912_}. 
On a more fundamental side, NMQSD has been studied in the context of continuous measurement theory \cite{Di08_080401_,WiGa08_140401_} and spontaneous  wavefunction localization \cite{BaFe09_050403_}.   
Other stochastic approaches, with various levels of applicability have been suggested \cite{StGr02_170407_,Sh04_5053_,PiMaH08_180402_}.

Although the NMQSD approach is formally exact, it seemed numerically difficult to handle, because of the appearance of a functional derivative with respect to a
stochastic process.
Only a few exactly solvable models are known (see e.g.~\cite{StDiGi99_1801_,JiYu10_240403_,FeBa12_170404_,JiZhYo12_042106_}). 
In previous works we have replaced that functional derivative by an operator ansatz and dealt with it in the so called ZOFE approximation \cite{YuDiGi99_91_,RoEiWo09_058301_,RoStEi11_034902_,RiRoSt11_113034_} that allows for a very efficient numerical solution and agrees remarkably well with established results for a large number of problems. 
However, in certain cases this method is known to fail (see e.g.~\cite{RoStEi11_034902_,RiRoSt11_113034_}). 
In Ref.~\cite{GaWi02_052105_} a hierarchical approach is applied to the operator ansatz of the functional derivative.  
 Our new HOPS presented here is not based on the previously assumed ansatz, it is numerically exact, converges rapidly and offers a systematic way to check for convergence by increasing the number of equations taken into account. 
In addition, it offers the advantages of stochastic Schr\"odinger equations, e.g.\  one deals with pure states (and not large density matrices) and the calculation of independent realizations can be parallelized trivially.

In the following, we first state the form of the open system problem we are interested in.
After a brief review of the general NMQSD approach we illustrate our new method for the case of zero temperature and an exponential bath-correlation function.
We derive a linear as well as the corresponding non-linear set of equations.
The latter is numerically more efficient and conceptually more interesting in terms of a pure state interpretation \cite{Di08_080401_,WiGa08_140401_}.
An extension to finite temperature and more general bath correlation function is presented afterwards.
We demonstrate the power of HOPS by applying it to the Spin-Boson model, the calculation of absorption spectra of molecular aggregates and energy transfer in a photosynthetic pigment-protein complex.
We use units where $k_{\rm B}=\hbar=1$.

\ssection{The Open Quantum System}
Let us consider a system linearly coupled to a bath of harmonic oscillators.
The Hamiltonian is a sum
\begin{equation}
  \op{H}_\mathrm{tot} = \op{H}\otimes\unity + \unity\otimes\op{H}_\mathrm{B} + H_\mathrm{int}\end{equation}
of the system Hamiltonian $\op{H}$, the bath Hamiltonian
\begin{equation}
  H_\mathrm{B} = \sum_\lambda \omega_{\lambda}\adj{a}_\lambda \op{a}_\lambda
\end{equation}
and the interaction Hamiltonian
\begin{equation}
  \op{H}_\mathrm{int} = \sum_\lambda ( \cc{g}_\lambda \op{L}\otimes\adj{a}_\lambda + g_\lambda\adj{L}\otimes\op{a}_\lambda ).
\end{equation}
Here, $\op{L}$ is an operator in the system's Hilbert space and $\adj{a}_\lambda$ the creation operator of bath mode $\lambda$.
The interaction strength between system and that mode is quantified by the complex number $g_\lambda$.
In many important cases one has $L=L^{\dagger}$.
It is convenient to encode the frequency dependence of the interaction strength  by the so called spectral density
$J(\omega)=\sum_j |g_{j}|^2 \delta (\omega-\omega_{j})$.
The latter is related to the bath correlation function $\BathCor(\tau)$ by \cite{MaKue00__}
\begin{equation}
  \label{eq:BathCor}
  \BathCor(\tau)= \int_0^{\infty}\! d\omega\, J(\omega)\Big(\coth\big(\frac{ \omega}{2 \kB T}\big)\, \cos(\omega \tau)
   - i \sin(\omega \tau) \Big)
\end{equation}
where $T$ is the temperature. 
Note that $\alpha(-\tau) = \cc\alpha(\tau)$.

In the following we are interested only in the dynamics in the system Hilbert space and in particular the reduced density matrix obtained by tracing over the bath degrees of freedom.

\ssection{Non-Markovian Quantum State Diffusion}
For now let us consider  initial conditions $\ket{\Psi_0} = \ket{\psi_0}\otimes\ket{\vec0}$, where $\ket{\vec0}$ is the vacuum state for all $\op{a}_\lambda$ in the bath Hilbert space (zero temperature). 
The reduced density matrix is 
\begin{equation}
  \label{eq:rho(t)}
  \rho_t=\Tr_B\{\ket{\Psi_t}\bra{\Psi_t}\},
\end{equation}
where $\Tr_B$ denotes the partial trace over the bath degrees of freedom and  $\ket{\Psi_t}$  is the solution of the Schrödinger equation $i \partial_t\ket{\Psi_t} = \op{H}_\mathrm{tot}\ket{\Psi_t}$.

Using a coherent state representation of the bath degrees of freedom, the reduced density matrix can be obtained from an ensemble average over trajectories of (non-normalized) pure states $\ket{\psi_t(\cc z)}$ in the system Hilbert space via
\begin{equation}
  \label{eq:rho(t)_z}
  \rho_t =  \E \big\{\ket{\psi_t(z^*)}\bra{\psi_t(z^*)}\big\},
\end{equation}
where $z=z_t$ is a complex Gaussian stochastic process with mean $\E \big[ z_t \big]=0$ and correlations
$\E\big[ z_t z_s \big]=0$
and
$\E\big[ z(t)z^{*}(s) \big]=\BathCor(t-s)$.
The time evolution of the states $\ket{\psi_t( z^*)}$ is determined \cite{DiSt97_569_,DiGiSt98_1699_} by
\begin{equation}
  \partial_t\psi_t = -\ii\op{H}\psi_t + \op{L}\ZZ_t\psi_t - \adj{L}\intl{0}{t}{s} \alpha(t-s)\frac{\delta \psi_t}{\delta \ZZ_s}
  \label{eq:nmsse}
\end{equation}
with initial conditions $\psi_{t=0} = \psi_0$.

While Eq.~(\ref{eq:rho(t)_z}) with (\ref{eq:nmsse}) determine the reduced density operator exactly, in general it is unclear how to solve Eq.~(\ref{eq:nmsse}) due to the functional derivative $\frac{\delta \psi_t}{\delta \ZZ_s}$.

In previous works we replaced this expression by an operator acting in the system Hilbert space $\frac{\delta \psi_t}{\delta \ZZ_s}=O(t,s,\ZZ)\psi_t$.
For some special cases this operator can be determined exactly \cite{DiGiSt98_1699_,JiZhYo12_exact_nmqsd}.
However, in general, approximation schemes are necessary (e.g.\ the ZOFE approximation \cite{RoEiWo09_058301_,RiRoSt11_113034_}).
Here we will proceed differently, without any approximation.

\ssection{HOPS -- Hierarchy of pure states}
First  Eq.~(\ref{eq:nmsse}) is written as
\begin{equation}
  \label{eq:first_order}
  \partial_t\psi_t = -\ii\op{H}\psi_t + \op{L}\ZZ_t\psi_t - \adj{L}\psi^{(1)}_t,
\end{equation}
 with the  auxiliary pure state
\begin{equation}
  \psi^{(1)}_t := \intl{0}{t}{s} \alpha(t-s) \frac{\delta\psi_t}{\delta\ZZ_s}.
\end{equation}
We now construct a hierarchy of equations by first considering the time derivative of $\psi^{(1)}_t$.
Note that one can write  $\psi^{(1)}_t = \adjZZ_t\psi_t$
where \footnote{ The bounded integral domain $[0,t]$ in \eqref{eq:nmsse} arises due to the initial condition $\ket{\Psi_0} = \ket{\psi_0}\otimes\ket{\vec0}$, which translates to $\frac{\delta\ket{\psi_{t=0}}}{\delta\ZZ_s}=0$ for $s \in \Reals$.
    Hence $\ket{\psi_t}$ must be independent of the noise $\ZZ_s$ for $s > t$ and $s < 0$.
}
\begin{equation}
  \adjZZ_t
  =  \intl{-\infty}{\infty}{s} \alpha(t-s) \frac{\delta}{\delta\ZZ_s}.
  \label{eq:adjZZt}
\end{equation}
Then $\dot\psi^{(1)}_t = \partial_t \left(\adjZZ_t \psi_t \right) = \dot\adjZZ_t\psi_t + \adjZZ_t\dot\psi_t$.
Reversing the argument that led to~\eref{eq:adjZZt} allows us to write $\dot\adjZZ_t\psi_t=\int_0^t ds \dot\alpha(t-s) \frac{\delta\psi_t}{\delta z_s^*}$.

In order to illustrate the derivation of the hierarchy of equations most clearly, we first consider  a bath-correlation function of the form
\begin{equation}
  \alpha(\tau) = g \, \exp[ - w \tau] \ (\tau\ge 0)  \mbox{ and }  \alpha(\tau) = \cc\alpha(-\tau) \ (\tau<0)
  \label{eq:exp_correlation_fun}
\end{equation}
with $w=\gamma + \ii\Omega$.
As shown for example in \cite{MeTa99_non_markovian}, sums of such exponentials are well suited to approximately describe a large class of spectral densities and also finite temperature.
For such an exponential correlation function one has $\dot\adjZZ_t\psi_t = -w \adjZZ_t \psi_t$ and thus  obtains
\begin{align}
  \partial_t\psi^{(1)}_t 
  &= -w\adjZZ_t\psi_t  -  \ii\op{H}\adjZZ_t\psi_t  +  \op{L}\adjZZ_t\ZZ_t\psi_t  -  \adj{L}\adjZZ_t^2\psi_t \\
  &= \left( -\ii\op{H} - w + \op{L}\ZZ_t \right) \psi^{(1)}_t  +  \alpha(0)\op{L}\psi^{(0)}_t  -  \adj{L}\psi^{(2)}_t.
\end{align}
with $\psi^{(k)}_t := \adjZZ_t^k\psi_t$.
In the first equality we used \eqref{eq:nmsse} as well as the fact that $\adjZZ_t$ commutes with all system operators.
The second equality follows from the commutator relation $\left[ \adjZZ_t, \ZZ_s \right] = \alpha(t-s)$.
By considering the time-derivatives of $\psi^{(k)}_t$ one gets coupled stochastic  equations for an infinite hierarchy of pure states (HOPS)
\begin{align}
  \partial_t\psi^{(k)}_t
  &= \left( -\ii\op{H} - k w + \op{L}\ZZ_t \right) \psi^{(k)}_t \nonumber \\
  &+ k\alpha(0)\op{L}\psi^{(k-1)}_t  -  \adj{L}\psi^{(k+1)}_t,
  \label{eq:linear_hierarchy}
\end{align}
with $\psi^{(0)}_{t=0} = \psi_0$ and $\psi^{(k)}_{t=0} = 0$ for $k>0$.
Solving the infinite system Eq.~(\ref{eq:linear_hierarchy}) is equivalent to solving Eq.~(\ref{eq:nmsse}), with $\psi_t=\psi^{(k=0)}_t$.
This is our first important result.

Clearly, our HOPS approach Eq.~(\ref{eq:linear_hierarchy}) has a similar structure as hierarchical equations of motions in the density operator formalism \cite{Ta06_082001_}.

\ssection{Truncation}
In order to transform Eq.~(\ref{eq:linear_hierarchy}) into a practical scheme, we truncate the  hierarchy at finite order.
In the present work, we use the following \quotes{terminator}
\begin{equation}
\psi^{(k+1)}_t \approx \frac{\alpha(0)}{w}\op{L}\psi^{(k)}_t
\end{equation}
for some suitable $k$ large enough.
Such a truncation is motivated by similar considerations 
as in Ref.~\cite{Ta06_082001_}.
By inserting the \quotes{terminator} into \eqref{eq:linear_hierarchy}, we obtain a closed system of $k + 1$ coupled equations.
We remark that the use of this particular terminator is not essential.
We have also found a good performance using $\psi^{(k+1)}_t=0$ with an appropriate $k$.

\ssection{Non-linear evolution equation }
The statistical properties of the linear system \eqref{eq:linear_hierarchy} of trajectories can be improved further by importance sampling: 
The Monte-Carlo determination of the density operator according to Eq.~\eqref{eq:rho(t)_z} converges much faster, if the contributions of individual realizations $\psi_t(\cc z)$ are of the same order of magnitude.
We therefore transform \eqref{eq:rho(t)_z}  to an average over normalized states. 
This can be achieved with the help of a Girsanov transformation, converting the linear equation \eqref{eq:nmsse} to a nonlinear form \cite{DiGiSt98_1699_}.
Using this construction as starting point we find the following hierarchy
\begin{equation}
  \begin{split}
    \dot{\tilde\psi}^{(k)}_t
    &=  \left( -\ii\op{H} - k w + \left( \ZZ_{t} + \intl{0}{t}{s} \cc\alpha(t-s) \langle\adj{L}\rangle_s  \right) \op{L}  \right)\tilde\psi^{(k)}_t \\
    &+  k\alpha(0) \op{L}\tilde\psi^{(k-1)}_t
    -  \left( \adj{L} - \langle\adj{L}\rangle_t \right) \tilde\psi^{(k+1)}_t.
  \end{split}
  \label{eq:nonlinear_hierarchy}
\end{equation}
Here, $\langle\cdot\rangle_s$ denotes the normalized average over $\tilde\psi_s^{(0)}$.
The terminator is the same as in the linear case, i.e.\ $\tilde\psi^{(k+1)}=(\alpha(0)/w)L\tilde\psi^{(k)}$.
Finally, the average in Eq.~(\ref{eq:rho(t)_z}) can now be performed over the normalized states $\tilde\psi_t\equiv\tilde\psi^{(k=0)}_t / \vert \tilde\psi^{(k=0)}_t \vert$.

\ssection{Generalizations}
We now generalize the results of the previous section to  bath-correlation functions  of the form
\begin{equation}
  \alpha(\tau)=\sum_{j=1}^J g_j e^{-w_j \tau} \ \ {\rm for}\ \tau \ge 0
  \label{eq:multiexp}
\end{equation}
with $w_j=\gamma_j+i \Omega_j$.
It is convenient to define tuples $\vec{w}=(w_1,\dots, w_J)$, which we indicate using boldface symbols.
The stochastic process $\ZZ_t$ corresponding to correlation function (\ref{eq:multiexp}) can be written as a sum of $J$  processes $\ZZ_{j,t}$.
Consequently, for each process we introduce an index $k_j$ that refers to the order of the corresponding hierarchy with auxiliary states $\psi^{(\kk)}_t = \psi^{(k_1,\dots,k_J)}_t$.
The full hierarchy of linear equations then reads
\begin{equation}
  \begin{split}
    \partial_t\psi^{(\kk)}
    &=  \left( -\ii\op{H} -\kk\cdot\vec{w} + \sum_j \op{L}\ZZ_{j,t} \right)\psi^{(\kk)}_t \\
    &+  \sum_j k_j \alpha_j(0) \psi^{(\kk-\ee_j)} - \sum_j \adj{L} \psi^{(\kk+\ee_j)}_t,
  \end{split}
\end{equation}
where $\kk\cdot\vec{w} = \sum_{j=1}^J k_jw_j$ determines the truncation condition and $\ee_j$ denotes the $j$-th unit vector in $\Reals^J$.
Applying the same reasoning to the triangular truncation condition $\abs{\kk} := k_1+\dots+k_J = \K$ leads to the generalized terminator
\begin{align}
  \psi^{(\kk+\ee_j)}
  = \sum_i  \frac{(\kk + \ee_j)_i \alpha_i(0)}{(\kk+\ee_j)\cdot\vec{w}} \op{L} \psi^{(\kk +\ee_j - \ee_i)}_t.
  \label{eq:terminator_triang}
\end{align}
Once again, one has to insert \eqref{eq:terminator_triang} in the last level of the hierarchy with $\abs{\vec k} = \K$.
The corresponding non-linear equation can be derived as in the case of a single exponential.

Note that depending on the situation different truncation conditions might be more efficient.
Furthermore, one can also treat independent environments ($n=1,2, \dots$) with different coupling operators $L_n$ along the same lines as required for the quantum aggregates below.

\ssection{Finite temperature}
 The case $T\! >\! 0$ can be mapped to the zero temperature case using the \emph{thermofield method} doubling the number of processes required \cite{DiGiSt98_1699_,Yu04_062107_,Ritschel_finite_Temp}.
 Remarkably, a system with self-adjoint coupling operator (i.e.\ $L\!=\!\adj{L}$) admits a description in terms of the zero-temperature non-Markovian quantum state diffusion equation \eqref{eq:nmsse} by introducing a sum process with correlation \eqref{eq:BathCor}.
For numerical efficiency we express $\alpha(\tau)$ as a sum of exponentials using the \emph{Pad\'{e} decomposition} of the hyperbolic cotangent \cite{JournalChemicalPhysics.134.244106}. 
The integral in Eq.~(\ref{eq:BathCor}) is then solved using the Residue theorem yielding the sought-after decomposition \eqref{eq:multiexp} with complex prefactors $g_j$.

\ssection{Spin-boson model}
\begin{figure}[t]
  \begin{center}
    \includegraphics[width=\columnwidth]{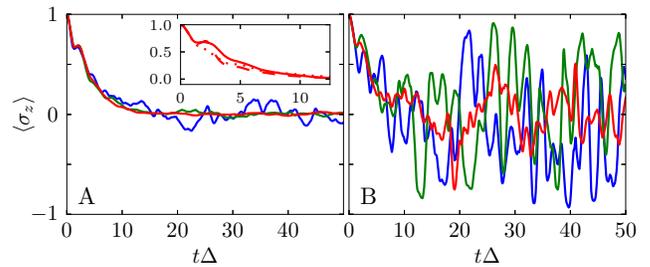}
  \end{center}
  \caption{
    (Color on-line) Dynamics of the spin-boson model.
    (A) Non-linear equation, (B) linear equation.
    In both cases $\Delta=1$, $\epsilon=0$ and the parameters of the spectral density are given by $g=2$ and $w=0.5 + 2\ii$. The blue, green, and red lines represent 100, 1000, and 10000 realizations respectively. The order of the hierarchy is $\mathcal{K}=8$. The inset in A shows the convergence (for 10000 realizations) with respect to the order of the hierarchy.
    Dotted, dashed, and solid line are orders one, two and four respectively.}
  \label{fig:spinboson}
\end{figure}
As a first example we consider the spin-boson model \cite{LeChDo87_1_}, where the system Hamiltonian is
$\op{H} = - \frac{1}{2}\Delta\op{\sigma}_x + \frac{1}{2}\epsilon\op{\sigma}_z$
and the coupling to the bath is mediated by
$\op{L}=\op{\sigma}_z$.
It is used to demonstrate the convergence of the method with respect to the truncation order of the hierarchy and with respect to the number of realizations.
In particular we show the superior convergence properties of the non-linear equation.
This can be clearly seen in Fig.~\ref{fig:spinboson}, where the dependence of the solution on the number of trajectories is shown for the non-linear (Fig.~\ref{fig:spinboson} A) and the linear equation  (Fig.~\ref{fig:spinboson} B).
While the non-linear solution already converges for 1 000 trajectories (and even for 100 trajectories is close to the converged solution), the linear equation shows large fluctuations even for 10 000 trajectories.
  Previous work \cite{VeAlGaSt05_temperature_nmsse} indicates that there is no significant difference between linear and non-linear variant in the weak coupling (Redfield) regime.
  However, in the strong coupling regime considered here, the non-linear version shows far superior convergence properties.
The inset displays how the solution of the non-linear equation converges with respect to the order of the hierarchy:
We observe converged results already at $\mathcal{K} = 4$.

\ssection{The quantum aggregate}
As an example for a more challenging setting we consider a generic system described by a Hamiltonian $H=\sum_n \epsilon_n \ket{n}\bra{n}+\sum_{nm}V_{nm}\ket{n}\bra{m}$, where $\ket{n}$ denotes a basis of the (small)  Hilbert space of the system.
In application to molecular aggregates, $\ket{n}$ denotes a localized electronic excitation at \quotes{site} $n$ of the system.
Each excitation couples to its own bath, that is  $\op{H}_\mathrm{int} = \sum_n \sum_\lambda ( \cc{g}_{n\lambda} \op{L}_n\otimes\adj{a}_{n\lambda} + g_{n\lambda}\adj{L}_n\otimes\op{a}_{n\lambda} )$ and $L_n=\ket{n}\bra{n}$.

\begin{figure}[tbp]
  \includegraphics[width=\columnwidth]{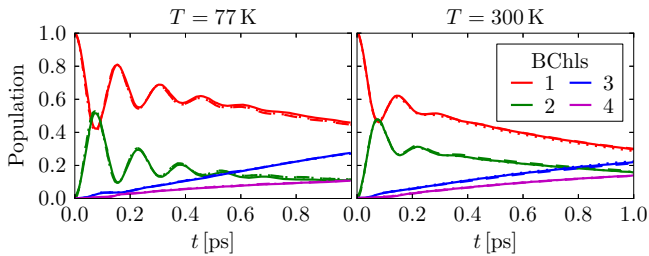}
  \caption{\label{fig:fmo_transfer}(Color on-line) Transfer of electronic excitation energy within the FMO complex. The parameters are taken from Ref.~\cite{IsFl09_17255_}. Solid line: result of  Ref.~\cite{IsFl09_17255_}, Dotted: HOPS first order, dashed: HOPS second order. }
\end{figure}
As an important application we consider transfer of electronic excitation within the photosynthetic FMO complex.
To demonstrate the accuracy of HOPS we compare with the numerical hierarchical equation of motion calculations of Ref.~\cite{IsFl09_17255_}.
As can be seen in Fig.~\ref{fig:fmo_transfer}, already the first order of HOPS agrees almost perfectly with the result of Ref.~\cite{IsFl09_17255_}.

\begin{figure}[t]
\includegraphics[width=8cm,height=3.cm]{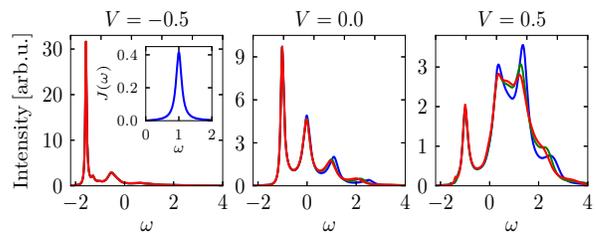}
  \caption{\label{fig:absorption}(Color on-line) Absorption of a linear chain of $N=7$ monomers with parallel transition dipoles for different values of the interaction $V$. The inset shows the spectral density used. The reorganization energy $E_r=\int_0^{\infty}\! dw J(w)/w $ is chosen as unit of energy. The temperature is $T=0.2$.
Colors indicate different orders of the hierarchy (blue: $\K=5$, green: $\K=6$, red: $\K=7$).   }
\end{figure}

Next we consider absorption of a linear aggregate:
Linear absorption can be calculated from the {\em linear} non-Markovian quantum state diffusion using only the single trajectory $\psi_t(z^{*}=0)$, i.e.\ no averaging over different realizations of the stochastic processes is needed \cite{RoStEi11_034902_,Ritschel_finite_Temp}.
We will now show that within our pure state hierarchy fast convergence of the optical spectra can be achieved.
To this end we employ the same model system as in Ref.~\cite{RoStEi11_034902_}, namely parallel transition dipoles and identical monomers.
In that case, the absorption strength for light with frequency $\nu$ can be calculated as \cite{RoStEi11_034902_}
\begin{equation}
  A(\nu) = \Re \intl{0}{\infty}{t} \exp[\ii\nu t] M(t),
\end{equation}
where $M$ is the correlation function
\begin{equation}
  M(t) = \mu^2 \sp{\psi_0(z^*)}{\psi_t(z^*)}|_{z^*=0}.
\end{equation}
Here, $\mu$ denotes the magnitude of the monomer's transition dipoles.
We have compared our HOPS calculations with numerically exact pseudo-mode calculations \cite{RoStEi11_034902_}.
For all cases considered we found perfect agreement with results of Ref.~\cite{RoStEi11_034902_} (not shown here).
In  Ref.~\cite{RoStEi11_034902_} only very short aggregates with  $N=2$ and $N=3$ (ignoring temperature) were considered, due to the huge numerical effort of the pseudo-mode approach.
With HOPS we are now able to study longer chains at finite temperature numerically exact.
As an example, in Fig.~\ref{fig:absorption} the absorption spectrum of a chain of 7 monomers is shown for the case of negative (left) and positive (right)  interaction $V$  together with the case of non interacting monomers (middle).
We have chosen a spiky spectral density (shown in the inset of panel a) and set the reorganization energy  $E_r=\int_0^{\infty}\! dw J(w)/w $ as unit of energy.
For the shown parameters $|V|=0.5$ and $T=0.2$ we are in the complicated case where all quantities are of the same order of magnitude and non-Markovian effects become clearly visible.
Note that the spectra converge faster at lower energies, so that already for small orders of the hierarchy one has a good description of the important low energy part of the spectrum.

\ssection{Conclusions and outlook}

The examples above demonstrate that our \emph{hierarchy of pure states} \textsc{HOPS} is very suitable to treat the dynamics of realistic open quantum systems covering strong coupling as well as highly non-Markovian regimes.
Based on a pure state representation \textsc{HOPS} is numerically efficient and converges fast towards the exact results.
While previous applications of non-Markovian quantum state diffusion rested on either analytically solvable models or approximation schemes, \textsc{HOPS} provides a numerically exact solution with a systematic control over potential errors.
Note that our formalism is not based on the unraveling of a given master equations as e.g.~non-Markovian Quantum Jumps \cite{BrPi09_50004_,ReChAs09_184102_}.
We obtain the reduced density operator directly from a closed system-environment model.
Since time dependent Hamiltonians can be included within the \textsc{HOPS} approach one can treat e.g.\ the interaction with an electromagnetic field as in femtosecond or 2D-spectroscopy.
Moreover, our quantum trajectory based formulation might help to shed light on quantum variants of fluctuation theorems \cite{CaTaHa09_fluctuation_theorems,EsMu06_fluctuation_theorems}.
We strongly believe that HOPS represents a fruitful approach to the study of dynamics of open quantum systems.

\vspace{0.5cm}

\begin{acknowledgments}
  We thank  
  Yoshitaka Tanimura for a helpful conversation about hierarchies,
  Ting Yu for his hospitality in Hoboken, 
 Frank Grossmann and Gerhard Ritschel for fruitful discussions 
  and John Briggs for many HOPS-inspired meetings.
\end{acknowledgments}


\begin{thebibliography}{10}
\providecommand{\url}[1]{\texttt{#1}}
\providecommand{\urlprefix}{URL }
\expandafter\ifx\csname urlstyle\endcsname\relax
  \providecommand{\doi}[1]{doi:\discretionary{}{}{}#1}\else
  \providecommand{\doi}{doi:\discretionary{}{}{}\begingroup
  \urlstyle{rm}\Url}\fi
\providecommand{\eprint}[2][]{\url{#2}}

\bibitem{We08__}
U.~Weiss; \emph{Quantum Dissipative Systems}; World Scientific Publishing
  Company; 3 edition edition (2008).

\bibitem{MaKue00__}
V.~May and O.~K{\"u}hn; \emph{{Charge and Energy Transfer Dynamics in Molecular
  Systems}}; WILEY-VCH (2000).

\bibitem{MaMa95_quapi}
N.~Makri and D.~E. Makarov; The Journal of Chemical Physics \textbf{102}
  (1995).

\bibitem{ThReHa00_quapi}
M.~Thorwart, P.~Reimann and P.~H\"anggi; Phys. Rev. E \textbf{62} 5808 (2000).

\bibitem{Ta06_082001_}
Y.~Tanimura; Journal of the Physical Society of Japan \textbf{75} 082001
  (2006).

\bibitem{KrKrRo11_2166_}
C.~Kreisbeck, T.~Kramer, M.~Rodr\'iguez and B.~Hein; Journal of Chemical Theory
  and Computation \textbf{7} 2166 (2011).

\bibitem{St96_25_}
W.~T. Strunz; Phys. Lett. A \textbf{224} 25 (1996).

\bibitem{Di96_309_}
L.~Di\'{o}si; Quantum and Semiclassical Optics: Journal of the European Optical
  Society Part B \textbf{8} 309 (1996).

\bibitem{DiSt97_569_}
L.~Di\'osi and W.~T. Strunz; Phys. Lett. A \textbf{235} 569 (1997).

\bibitem{DiGiSt98_1699_}
L.~Di\'osi, N.~Gisin and W.~T. Strunz; Phys. Rev. A \textbf{58} 1699 (1998).

\bibitem{RiRoSt11_113034_}
G.~Ritschel, J.~Roden, W.~T. Strunz and A.~Eisfeld; New Journal of Physics
  \textbf{13} 113034 (2011).

\bibitem{RiRoSt11_2912_}
G.~Ritschel, J.~Roden, W.~T. Strunz, A.~Aspuru-Guzik and A.~Eisfeld; The
  Journal of Physical Chemistry Letters \textbf{2} 2912 (2011).

\bibitem{Di08_080401_}
L.~Di\'osi; Phys. Rev. Lett. \textbf{100} 080401 (2008).

\bibitem{WiGa08_140401_}
H.~M. Wiseman and J.~M. Gambetta; Phys. Rev. Lett. \textbf{101} 140401 (2008).

\bibitem{BaFe09_050403_}
A.~Bassi and L.~Ferialdi; Phys. Rev. Lett. \textbf{103} 050403 (2009).

\bibitem{StGr02_170407_}
J.~T. Stockburger and H.~Grabert; Phys. Rev. Lett. \textbf{88} 170407 (2002).

\bibitem{Sh04_5053_}
J.~Shao; The Journal of Chemical Physics \textbf{120} (2004).

\bibitem{PiMaH08_180402_}
J.~Piilo, S.~Maniscalco, K.~H\"ark\"onen and K.-A. Suominen; Phys. Rev. Lett.
  \textbf{100} 180402 (2008).

\bibitem{StDiGi99_1801_}
W.~T. Strunz, L.~Di\'osi and N.~Gisin; Phys. Rev. Lett. \textbf{82} 1801
  (1999).

\bibitem{JiYu10_240403_}
J.~Jing and T.~Yu; Phys. Rev. Lett. \textbf{105} 240403 (2010).

\bibitem{FeBa12_170404_}
L.~Ferialdi and A.~Bassi; Phys. Rev. Lett. \textbf{108} 170404 (2012).

\bibitem{JiZhYo12_042106_}
J.~Jing, X.~Zhao, J.~Q. You and T.~Yu; Phys. Rev. A \textbf{85} 042106 (2012).

\bibitem{YuDiGi99_91_}
T.~Yu, L.~Diosi, N.~Gisin and W.~T. Strunz; Phys. Rev. A \textbf{60} 91 (1999).

\bibitem{RoEiWo09_058301_}
J.~Roden, A.~Eisfeld, W.~Wolff and W.~T. Strunz; Phys. Rev. Lett. \textbf{103}
  058301 (2009).

\bibitem{RoStEi11_034902_}
J.~Roden, W.~T. Strunz and A.~Eisfeld; J. Chem. Phys. \textbf{134} 034902
  (2011).

\bibitem{GaWi02_052105_}
J.~Gambetta and H.~M. Wiseman; Phys. Rev. A \textbf{66} 052105 (2002).

\bibitem{JiZhYo12_exact_nmqsd}
J.~Jing, X.~Zhao, J.~Q. You and T.~Yu; Phys. Rev. A \textbf{85} 042106 (2012).

\bibitem{MeTa99_non_markovian}
C.~Meier and D.~J. Tannor; The Journal of chemical physics \textbf{111} 3365
  (1999).

\bibitem{Yu04_062107_}
T.~Yu; Phys. Rev. A \textbf{69} 062107 (2004).

\bibitem{Ritschel_finite_Temp}
G.~Ritschel, D.~S\"u\ss, W.~T. Strunz and A.~Eisfeld; In preparation  (2014).

\bibitem{JournalChemicalPhysics.134.244106}
J.~Hu, M.~Luo, F.~Jiang, R.-X. Xu and Y.~Yan; The Journal of Chemical Physics
  \textbf{134} 244106 (2011).

\bibitem{LeChDo87_1_}
A.~J. Leggett, S.~Chakravarty, A.~T. Dorsey, M.~P.~A. Fisher, A.~Garg and
  W.~Zwerger; Rev. Mod. Phys. \textbf{59} 1 (1987).

\bibitem{VeAlGaSt05_temperature_nmsse}
I.~de~Vega, D.~Alonso, P.~Gaspard and W.~T. Strunz; The Journal of Chemical
  Physics \textbf{122} 124106 (2005).

\bibitem{IsFl09_17255_}
A.~Ishizaki and G.~R. Fleming; PNAS \textbf{106} 17255 (2009).

\bibitem{BrPi09_50004_}
H.-P. Breuer and J.~Piilo; Euro. Phys. Lett. \textbf{85} 50004 (2009).

\bibitem{ReChAs09_184102_}
P.~Rebentrost, R.~Chakraborty and A.~Aspuru-Guzik; J. Chem. Phys. \textbf{131}
  184102 (2009).

\bibitem{CaTaHa09_fluctuation_theorems}
M.~Campisi, P.~Talkner and P.~H\"anggi; Phys. Rev. Lett. \textbf{102} 210401
  (2009).

\bibitem{EsMu06_fluctuation_theorems}
M.~Esposito and S.~Mukamel; Phys. Rev. E \textbf{73} 046129 (2006).

\end{thebibliography}
\end{document}